\documentclass[fleqn,usenatbib]{mnras}

\usepackage{newtxtext,newtxmath}

\usepackage[T1]{fontenc}

\DeclareRobustCommand{\VAN}[3]{#2}
\let\VANthebibliography\thebibliography
\def\thebibliography{\DeclareRobustCommand{\VAN}[3]{##3}\VANthebibliography}

\usepackage{graphicx}	
\usepackage{amsmath}	

\newcommand{\objafull}{DES~J024703.24$-$010032.0}
\newcommand{\obja}{DES~J0247$-$0100}
\newcommand{\objbfull}{DES~J024944.66$-$000036.8}
\newcommand{\objb}{DES~J0249$-$0000}
\newcommand{\objcfull}{DES~J025214.67$-$002813.7}
\newcommand{\objc}{DES~J0252$-$0028}

\title[VLA imaging of Periodic Quasars]{Very Large Array imaging rules out precessing radio jets in three DES--SDSS-selected candidate periodic quasars}

\author[Chen, Liu, Liao, Guo]{
Yu-Ching Chen,$^{1,2,3}$\thanks{E-mail: ycchen@illinois.edu (YCC); xinliuxl@illinois.edu (XL)}
Xin Liu,$^{1,2}$
Wei-Ting Liao,$^{1,2}$
and Hengxiao Guo$^{4}$
\\
$^{1}$Department of Astronomy, University of Illinois at Urbana-Champaign, 1002 West Green Street, Urbana, IL 61801, USA\\
$^{2}$National Center for Supercomputing Applications, 1205 West Clark Street, Urbana, IL 61801, USA\\
$^{3}$Center for AstroPhysical Surveys, National Center for Supercomputing Applications, Urbana, IL, 61801, USA\\
$^{4}$Department of Physics and Astronomy, 4129 Frederick Reines Hall, University of California, Irvine, CA 92697-4575, USA\\
}

\date{Accepted XXX. Received YYY; in original form ZZZ}

\pubyear{2021}

\begin{document}
\label{firstpage}
\pagerange{\pageref{firstpage}--\pageref{lastpage}}
\maketitle

\begin{abstract}
Periodic quasars have been suggested as candidates for hosting binary supermassive black holes (SMBHs), although alternative scenarios remain possible to explain the optical light curve periodicity. To test the alternative hypothesis of precessing radio jet, we present deep 6 GHz radio imaging conducted with NSF's Karl G. Jansky Very Large Array (VLA) in its C configuration for the three candidate periodic quasars, \objafull , \objbfull , and \objcfull . Our targets were selected based on their optical variability using 20-yr long multi-color light curves from the Dark Energy Survey (DES) and the Sloan Digital Sky Survey (SDSS). The new VLA observations show that all three periodic quasars are radio-quiet with the radio loudness parameters measured to be $R\equiv f_{6\,{\rm cm}}/f_{{\rm 2500}}$ of $\lesssim$1.0--1.5 and the $k$-corrected luminosities $\nu L_\nu$[6 GHz] of $\lesssim$5--21 $\times$ 10$^{39}$ erg s$^{-1}$. They are in stark contrast to previously known periodic quasars proposed as binary SMBH candidates such as the blazar OJ 287 and PG1302$-$102. Our results rule out optical emission contributed from precessing radio jets as the origin of the optical periodicity in the three DES--SDSS-selected candidate periodic quasars. Future continued optical monitoring and complementary multi-wavelength observations are still needed to further test the binary SMBH hypothesis as well as other competing scenarios to explain the optical periodicity. 
\end{abstract}




\section{Introduction}
Periodic quasars have been suggested as close binary supermassive black holes (SMBHs) – the progenitors of SMBH mergers \citep[e.g.,][]{haiman09,loeb10,Tanaka2013,Farris2014,Shi2015,Kelley2019,Duffell2020,Serafinelli2020,Xin2020b,Xin2021}, which are of interest as sources of low frequency gravitational waves \citep[e.g.,][]{Burke-Spolaor2013,Mingarelli2017,Middleton2018,Zhu2019,Nguyen2020,Sesana2021,Taylor2021} and possibly for the formation of SMBH seeds at high redshift \citep[e.g.,][]{Kroupa2020}. Binary SMBHs are expected to form in galaxy mergers \citep{begelman80,yu02,DEGN}, given that most massive galaxies harbor SMBHs \citep{kormendy95,KormendyHo2013,Heckman2014}.  While ${\sim}160$ periodic quasars have been proposed as close binary SMBH candidates \citep[e.g.,][]{valtonen08,Graham2015,Charisi2016,Zheng2016,Li2016,Li2019,Liutt2019,chenyc2020,liao2020}, no confirmed case is known at sub-milliparsec scales, i.e., separations close enough to be in the gravitational wave regime.

Significant periodicities were recently discovered in the optical light curves of five $z\sim2$ quasars \citep{chenyc2020,liao2020}. The systematic search combined new, highly sensitive light curves from the Dark Energy Survey Supernova (DES-SN) fields \citep{Kessler2015,Tie2017} with archival data from the Sloan Digital Sky Survey (SDSS) Stripe 82 \citep{ivezic07}. Though periodic quasar candidates are usually linked to binary SMBHs, the true physical mechanism driving the periodicity is still unclear. Doppler beaming effect of a accretion disk \citep{DOrazio2015,Charisi2018} predicts sinusoidal signals with wavelength dependence, whereas circumbinary accretion disk variability seen in the hydrodynamical simulations \citep[e.g.,][]{Farris2014,Duffell2020} often shows bursty light curves. Jet precession is also proposed to explain some characteristics of the radio-loud periodic quasars such as OJ287, 3C120, PG1553+113 \citep{Katz1997, Caproni2004, Caproni2017}. The accretion disk or/and the jet of a secondary black hole that is inclined to the orbital plane of a binary system will undergo relativistic precession and produce periodic signal due to variation in the viewing angle \citep{Lehto1996}. Based on the frequency dependence of the variability amplitudes and the shapes of the light curves, it is likely that the periodicities of the five candidates in \citet{chenyc2020} are originated from hydrodynamic circumbinary accretion disk variability instead of Doppler beaming. 


While a binary SMBH is perhaps the most plausible explanation, there are alternative interpretations for the observed optical periodicity with a single rotating SMBH such as precession of a warped accretion disk or/and a radio jet, or instability in accretion flow \citep[e.g.,][]{Stella1998,Rieger2004,Vaughan2005a,Caproni2007,Ingram2009,Tchekhovskoy2011,McKinney2012,King2013,Tremaine2014}. One popular model is Lense–Thirring precession of a geometrically thick, accretion disk near the central black hole. Due to misalignment of the BH spin axis with the angular momentum of accretion disk, Lense–Thirring effect \citep{Bardeen1975} predicts precession of a accretion disk as well as a radio jet if it presents.  Other possibilities are usually associated with instability or oscillation of acrretion flow often seen in the three-dimensional magnetohydrodynamic simulations \citep{Tchekhovskoy2011,McKinney2012}

In this work, we test the leading hypothesis -- radio jet precession for three candidate periodic quasars, \objafull\ (hereafter \obja\ for short) and \objbfull\ (hereafter \objb\ for short) discovered by \citet{chenyc2020} and \objcfull\ (hereafter \objc\ for short) discovered by \citet{liao2020} by conducting new radio observations with NSF's Karl G. Jansky Very Large Array (VLA). 


This paper is organised as follows. In \S\ref{sec:observations}, we describe our new VLA observations, archival DES imaging data, SDSS spectra and other supplementary data. In \S\ref{sec:results}, we describe the results of the radio analysis. In \S\ref{sec:discussions}, we discuss the origin of the radio emission, followed by the comparison with the previous periodic quasars and the implications of the results for the jet precession scenario. Finally, we summarize our results and conclude in \S\ref{sec:conclusions}. A concordance $\Lambda$CDM cosmology with $\Omega_m = 0.3$, $\Omega_{\Lambda} = 0.7$, and $H_{0}=70$ km s$^{-1}$ Mpc$^{-1}$ is assumed throughout.

\begin{table*}
 \caption{Properties of the DES--SDSS-selected candidate periodic quasars and the detail of the VLA observations. Listed from left to right are the source name in J2000 coordinates, spectroscopic redshift from the SDSS quasar catalog \citep{Paris2018}, optical co-added PSF magnitude in $r$ band, date of VLA observation, VLA array configuration, center frequency, total bandwidth, synthesized beam size, position angle of synthesized beam, and 1-$\sigma$ rms noise level. Synthesized beam size, position angle and 1-$\sigma$ rms noise level are calculated using nature weighting. 
 }
 \label{tab:obs}
 \begin{tabular}{cccccccccc}
  \hline\hline
  Target Name & Redshift & $m_r$ &  Obs. date & Config. & Frequency & Bandwidth & Beam Size & P.A. & r.m.s.\\
  &  & [mag] & [UTC] &  & [GHz] & [GHz] & [\arcsec$\times$\arcsec] & [deg] & [$\mu$Jy beam$^{-1}$] \\
  \hline
  \objafull & 2.525 & 20.33 & 2020-06-05 & C & 6.0 & 4.0 & 6.0$\times$3.4 & $-$38.8 & 3.0 \\
  \objbfull & 1.295 & 21.04 & 2020-05-10 & C & 6.0 & 4.0 & 7.6$\times$3.9 & 46.7 & 3.0\\
  \objcfull & 1.530 & 20.60 & 2020-05-05 & C & 6.0 & 4.0 & 5.1$\times$3.6 & $-$22.8 & 3.0 \\
  \hline
  \end{tabular}
\end{table*}

\section{Observations and Data Analysis} \label{sec:observations}

\subsection{New VLA Radio Continuum Observations}

We observed three candidate periodic quasars \obja , \objb , and \objc\ with the VLA in its C configuration at the central frequency of 6.0 GHz on May 5, May 10, and June 5 in 2020 under Program ID 20A-457 (PI Liu). 3C 48 was used as the flux and bandpasss calibrator and J0239$-$0234 was used as the gain calibrator. The on-source observing time was 53 mins per target. The images were calibrated and reduced through the standard VLA Calibration Pipeline version 5.6.2\footnote{\url{https://science.nrao.edu/facilities/vla/data-processing/pipeline/CIPL_56}} in the Common Astronomy Software Applications package \citep[CASA,][]{McMullin2007}. With a total bandwidth of 4 GHz and nature weighting, the achieved 1-$\sigma$ rms noise level of the cleaned images is $\sim$3.0 $\mu$Jy/beam. \autoref{tab:obs} lists details of the VLA observations.


\subsection{Optical Continuum Flux Densities from Archival SDSS Spectra}

To calculate the radio loudness, we need to estimate the optical continuum flux densities at the rest-frame 2500 {\rm \AA} that are measured at the same time as those of the radio continuum flux densities. We calculate the optical continuum flux densities of the three targets based on the archival optical spectra from the SDSS corrected for variability using the long-term optical light curve in the optical band that contains the rest-frame 2500 {\rm \AA}. 

The three quasars were observed with the BOSS spectrograph by the SDSS-III/BOSS survey \citep{Dawson2013} or the SDSS-IV/eBOSS survey \citep{Dawson2016}. The BOSS spectra cover observed 3650--10400 {\rm \AA} with a spectral resolution of $R=$1850--2200. We fit the SDSS spectra following the procedures described in \cite{Shen2012} using the code PyQSOFit \citep{Guo2018,Shen2019}. The spectral model consists of a power-law continuum, a pseudo-continuum constructed from the Fe II emission templates, and single or multiple Gaussian components for the narrow and broad emission lines. 

\subsection{Archival Optical Imaging and New Optical Light Curves}

Our targets have publicly available optical images from the Dark Energy Survey \citep[DES;][]{Flaugher2005,DES,Diehl2019}. DES is a 6-yr optical imaging survey of the southern sky in five filters ($grizY$). DES has a median seeing of 1\arcsec~ and a nominal single-epoch depth down to $r\sim$23.34 \citep{Morganson2018}.
The optical light curves of our targets include photometric measurements from various optical surveys including DES, SDSS \citep{SDSSDR5,ivezic07}, the Catalina Real Time Transient Survey \citep[CRTS;][]{Drake2009}, the Palomar Transient Factory \citep[PTF;][]{PTF}, the Panoramic Survey Telescope and Rapid Response System \citep{Chambers2016,Flewelling2016}, and the Zwicky Transient Facility \citep{ZTF_DR1}. 
To correct for the variability effect in the radio loudness calculation, we fit a sinusoidal model to the optical light curves to estimate the optical continuum flux densities at the same time as those of the radio observations.

\section{Results}
\label{sec:results}

\begin{figure*}
    \centering
    \includegraphics[width=0.25\linewidth]{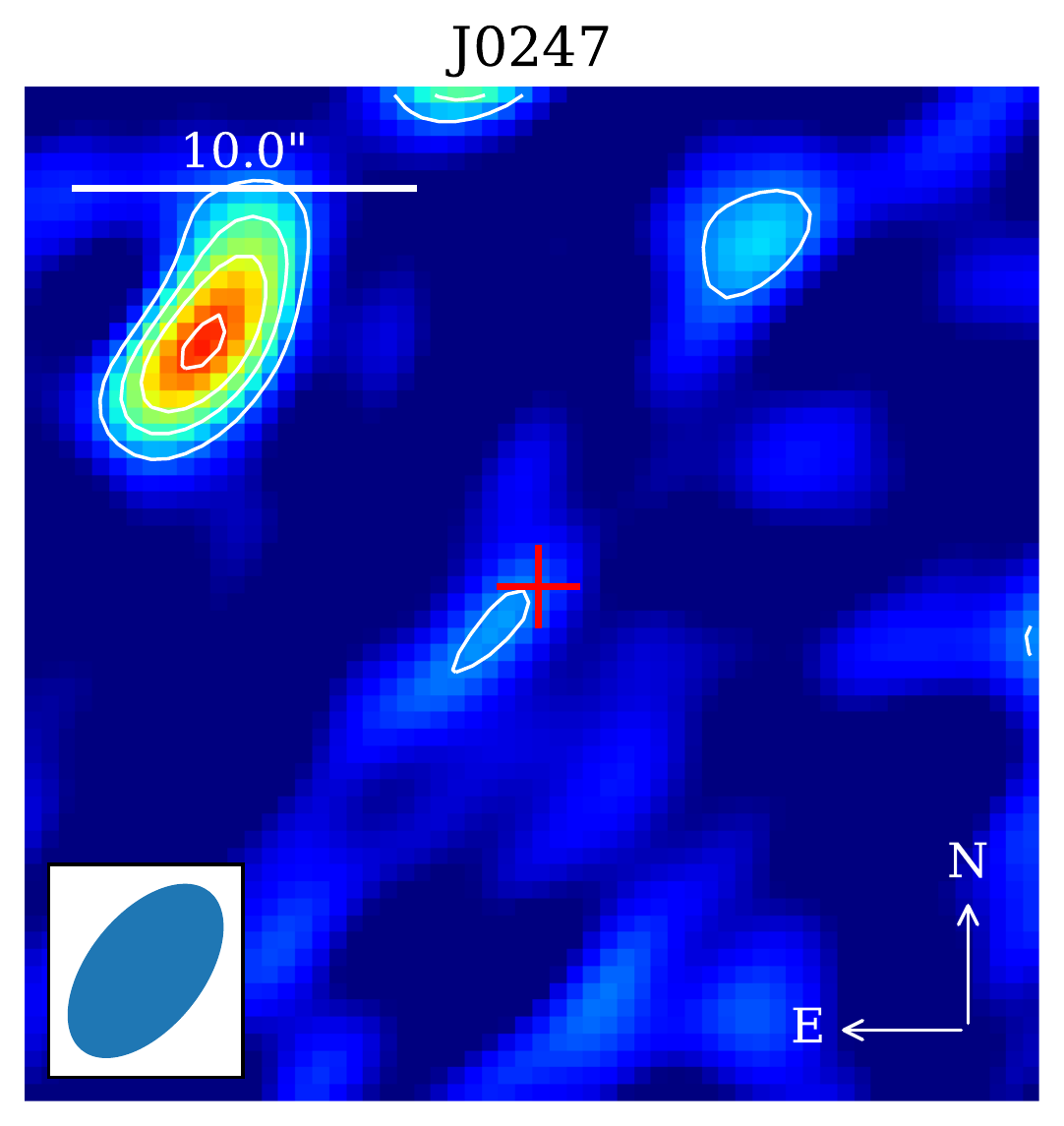}
    \includegraphics[width=0.25\linewidth]{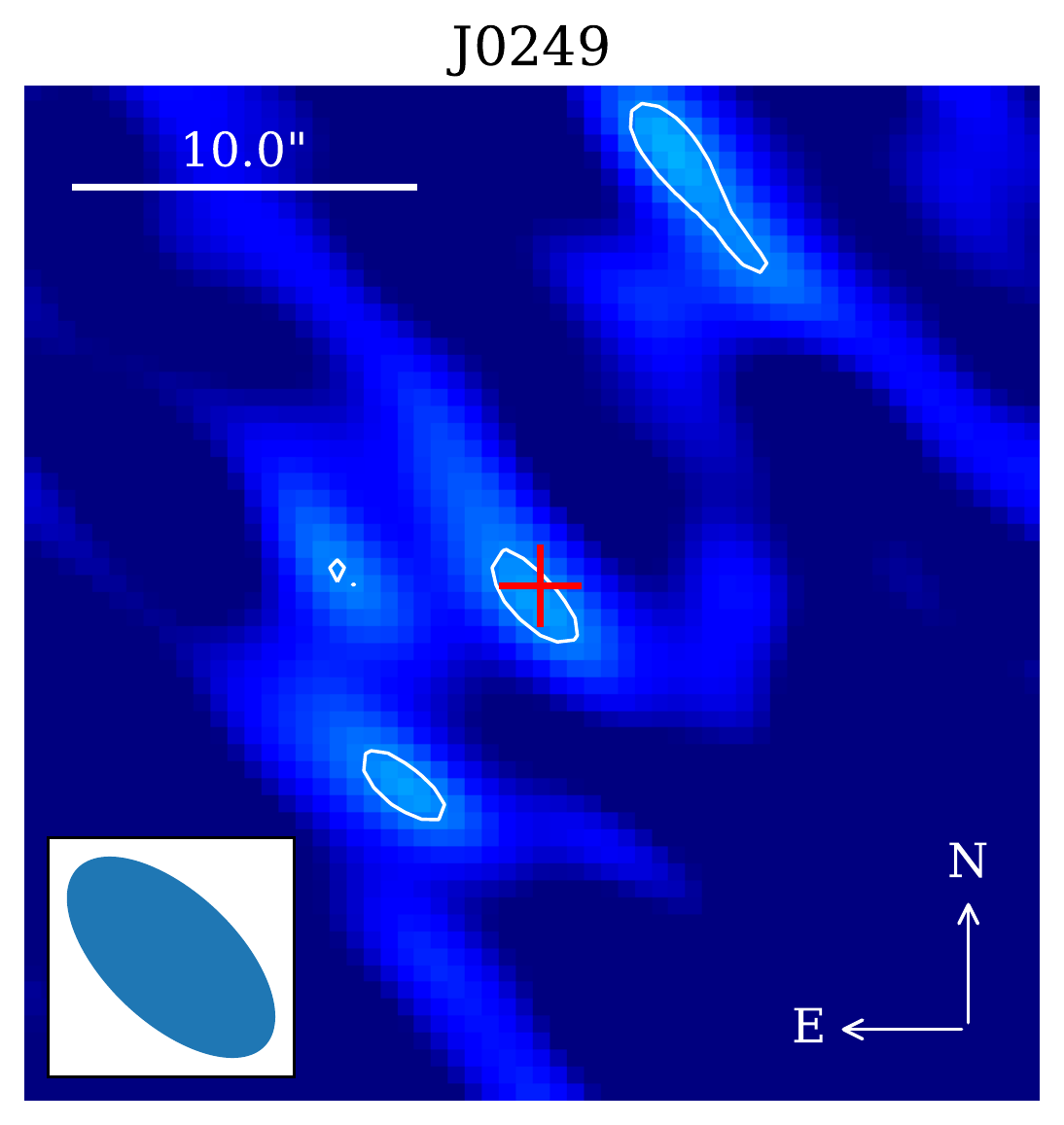}
    \includegraphics[width=0.25\linewidth]{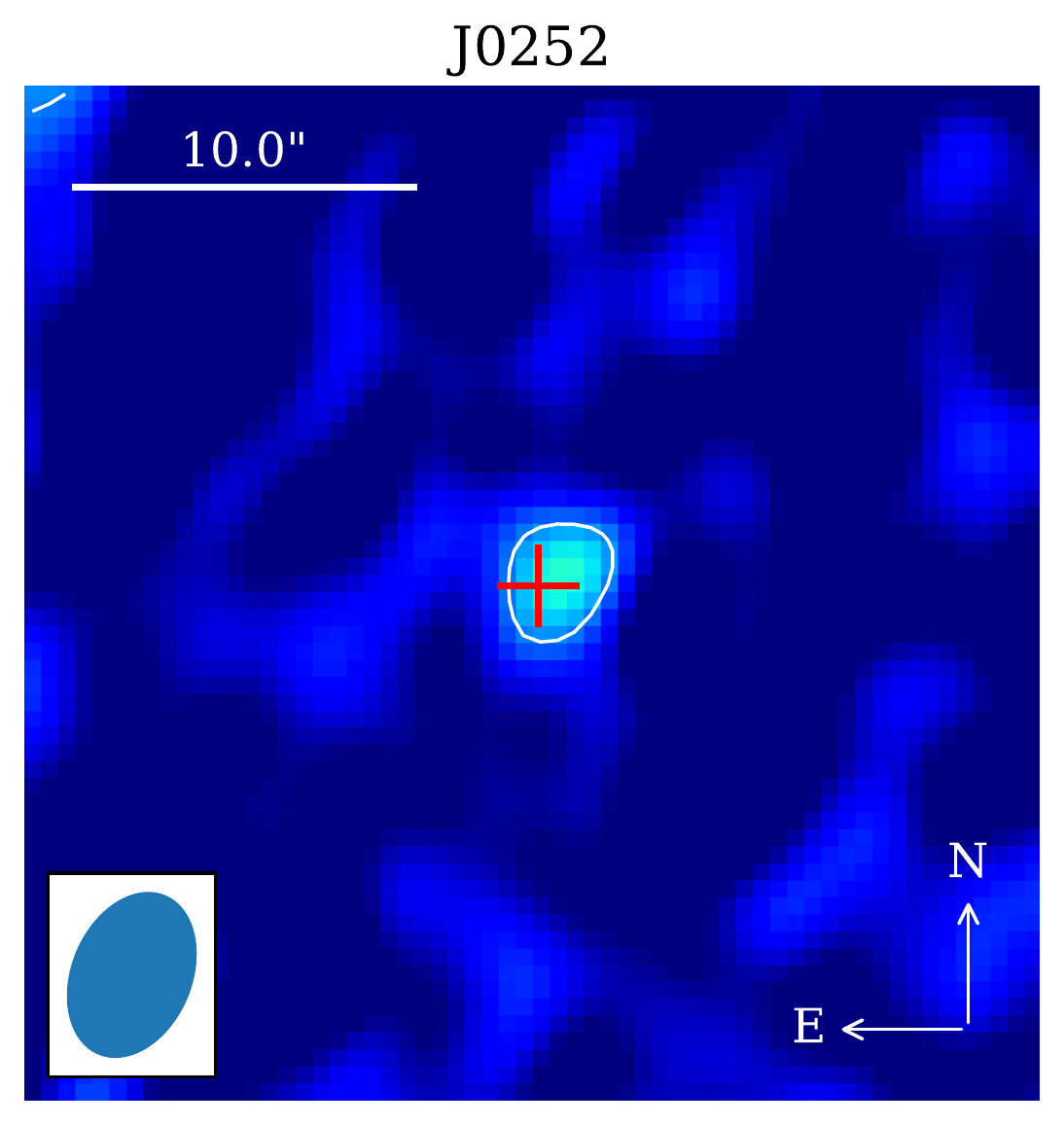}
    \caption{Deep 6 GHz VLA images of the three DES--SDSS-selected candidate periodic quasars ({\obja}, {\objb}, and {\objc}). 
    {\objc} shows a $\sim$5$\sigma$ detection, whereas {\obja} and {\objb} are not clearly detected, whose $\sim$3$\sigma$ detections are possible structured noises from the clean processes.} The contour levels are 3$\sigma$, 5$\sigma$, 7$\sigma$, 10$\sigma$, 15$\sigma$, and 20$\sigma$, where the 1$\sigma$ rms noise level is 3.0 $\mu$Jy/beam. The image field of view for each target is 30\arcsec$\times$30\arcsec. The red cross marks the peak flux position of the optical counterpart shown in \autoref{fig:des_images}. The synthesized beam is shown in the bottom left corner.
    \label{fig:vla_images}
\end{figure*}

\begin{figure*}
    \centering
    \includegraphics[width=0.25\linewidth]{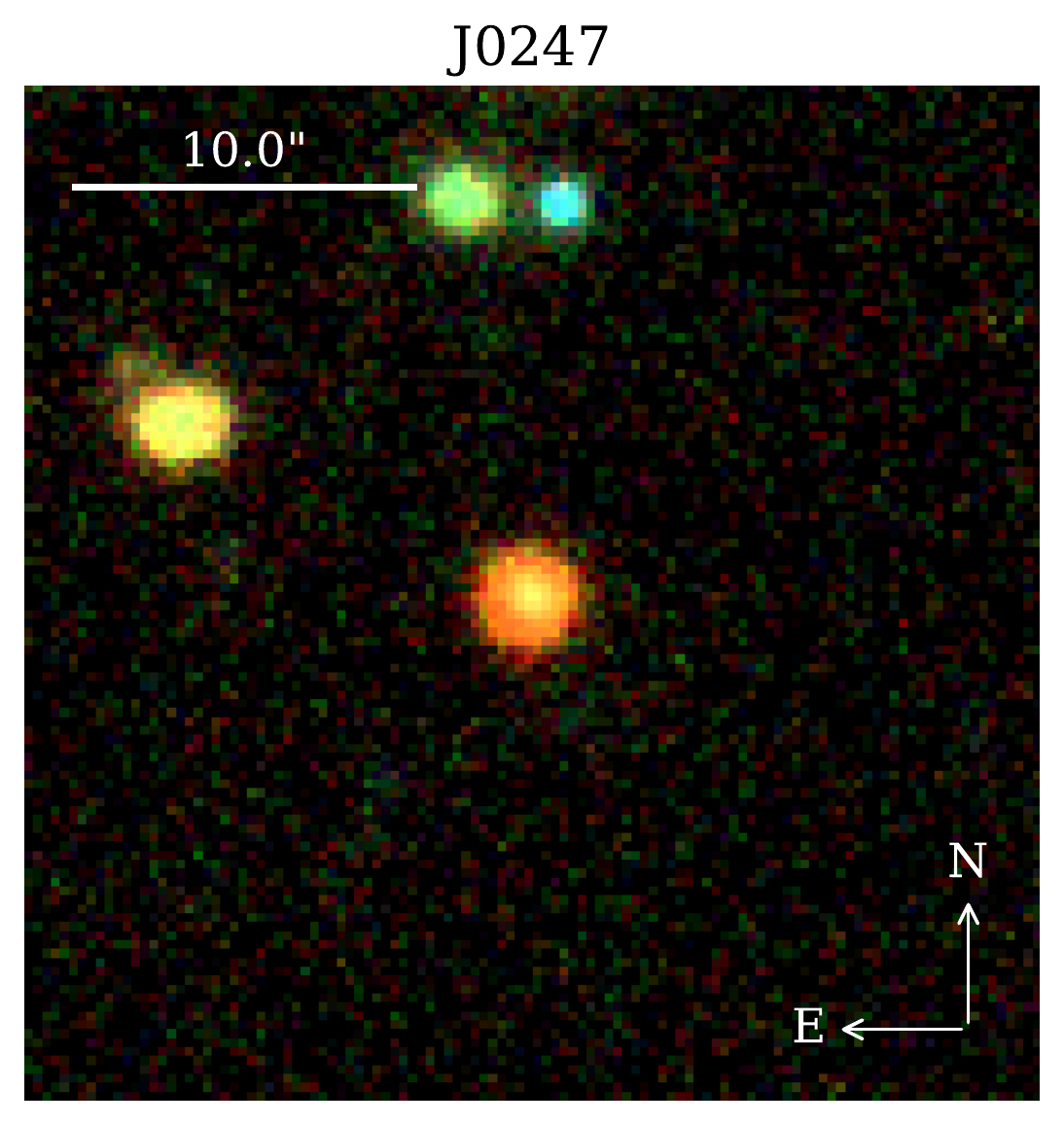}
    \includegraphics[width=0.25\linewidth]{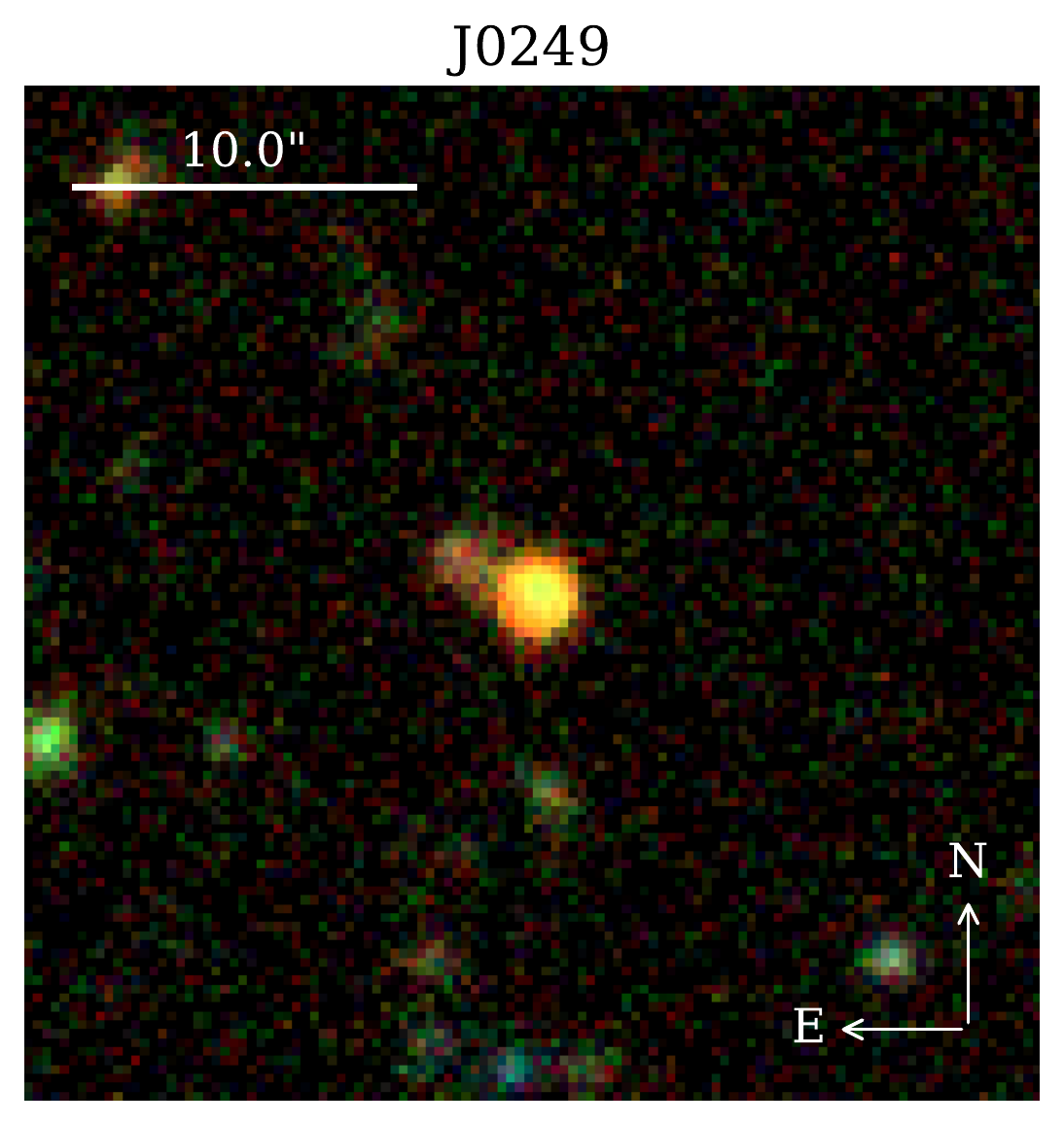}
    \includegraphics[width=0.25\linewidth]{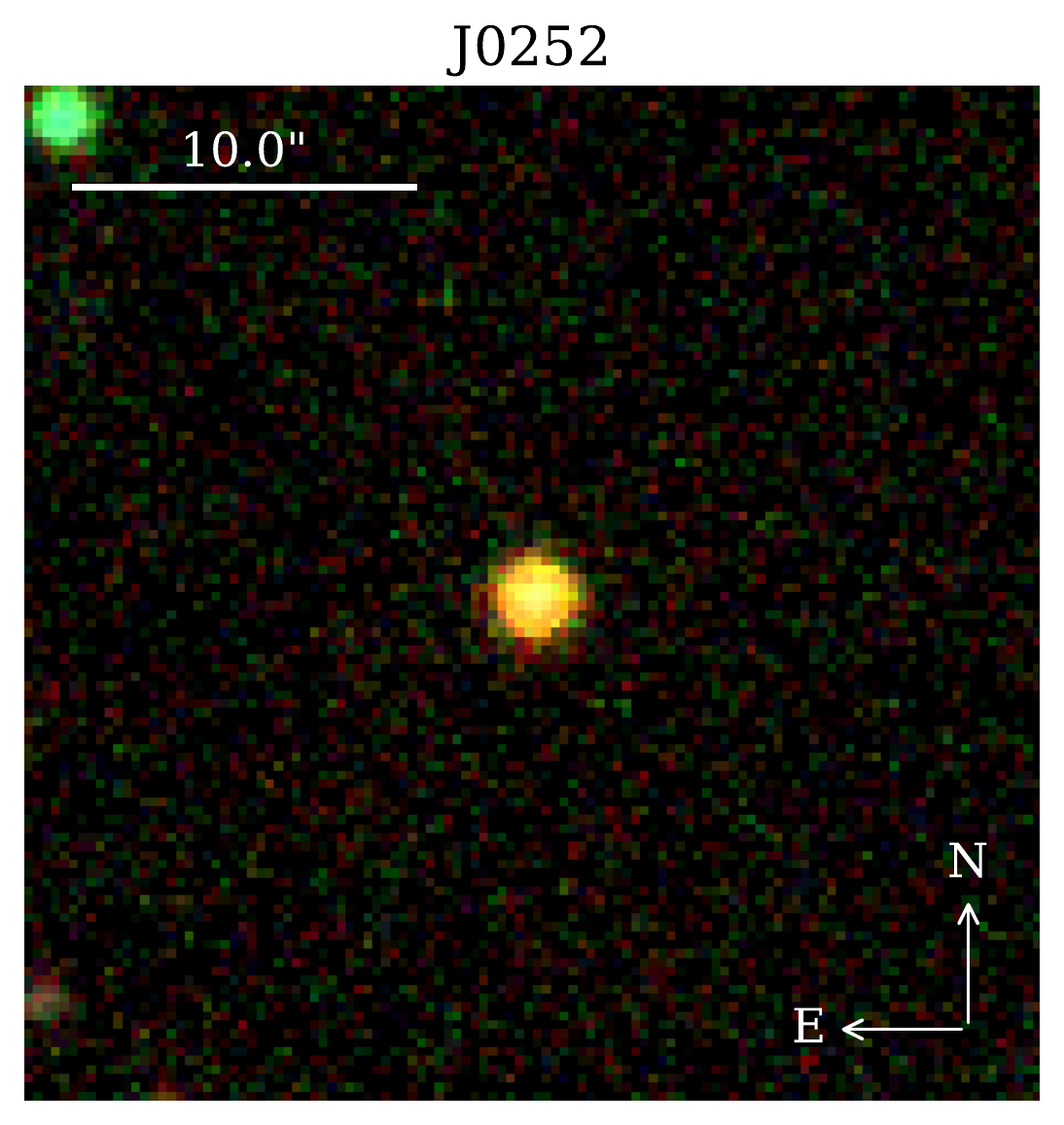}
    \caption{DES $gri$ color composite images of the three DES--SDSS-selected candidate periodic quasars ({\obja}, {\objb}, and {\objc}).The image field of view for each target is 30\arcsec$\times$30\arcsec.}
    \label{fig:des_images}
\end{figure*}

\subsection{VLA Radio Images and Comparison to DES Optical Images}

\autoref{fig:vla_images} shows the 6 GHz VLA continuum images of the three DES--SDSS-selected candidate periodic quasars using nature weighting. 
\objc\ shows a $\sim$5$\sigma$ detection, whereas \obja\ and \objb\ are not clearly detected, whose $\sim$3$\sigma$ detections are likely the structured noises from the clean processes. There are no ($>$3$\sigma$) extended radio jet structures associated with the central cores. We fit the data with a 2-dimensional Gaussian function to obtain the integrated flux density for \obja\ and use 3$\sigma$ upper limits for \obja\ and \objb. The integrated 6 GHz radio flux densities are $<$9, $<$9 and $\sim$13$\pm$3 $\mu$Jy for {\obja}, {\objb}  and {\objc}, respectively (\autoref{tab:info}). 

For comparison, \autoref{fig:des_images} shows the DES $gri$ color composite images of the three targets. The optical morphology are consistent with the Point Spread Function (PSF) except for a faint extended emission 2{\arcsec} N-E of the central source in {\objb}, although no extended radio emission is detected at the same position.

\subsection{Radio Luminosities and Radio Loudness Parameters} 

\label{sec:results-RL}


The $k$-corrected intrinsic radio luminosity at rest-frame 6 GHz can be estimated by
\begin{equation}
    \nu L_{\nu} = \frac{4\pi D_L^2 \nu f_{\nu} }{(1+z)^{1+\alpha}},
\end{equation}
where $\nu =$ 6 GHz, $D_L$ is the luminosity distance, $z$ is the redshift, $\alpha$ is the spectral index, and $f_{\nu}$ is the observed flux density . Assuming a radio spectral index of $\alpha=-0.8$ \citep{Kimball2008}, the radio luminosities at rest-frame 6 GHz are estimated as $<2.1\times10^{40}$, $<4.5\times10^{39}$ and ${\sim}9.7\times10^{39}$ erg s$^{-1}$ for {\obja}, {\objb} and {\objc}, respectively.

\begin{figure}
    \centering
    \includegraphics[width=\linewidth]{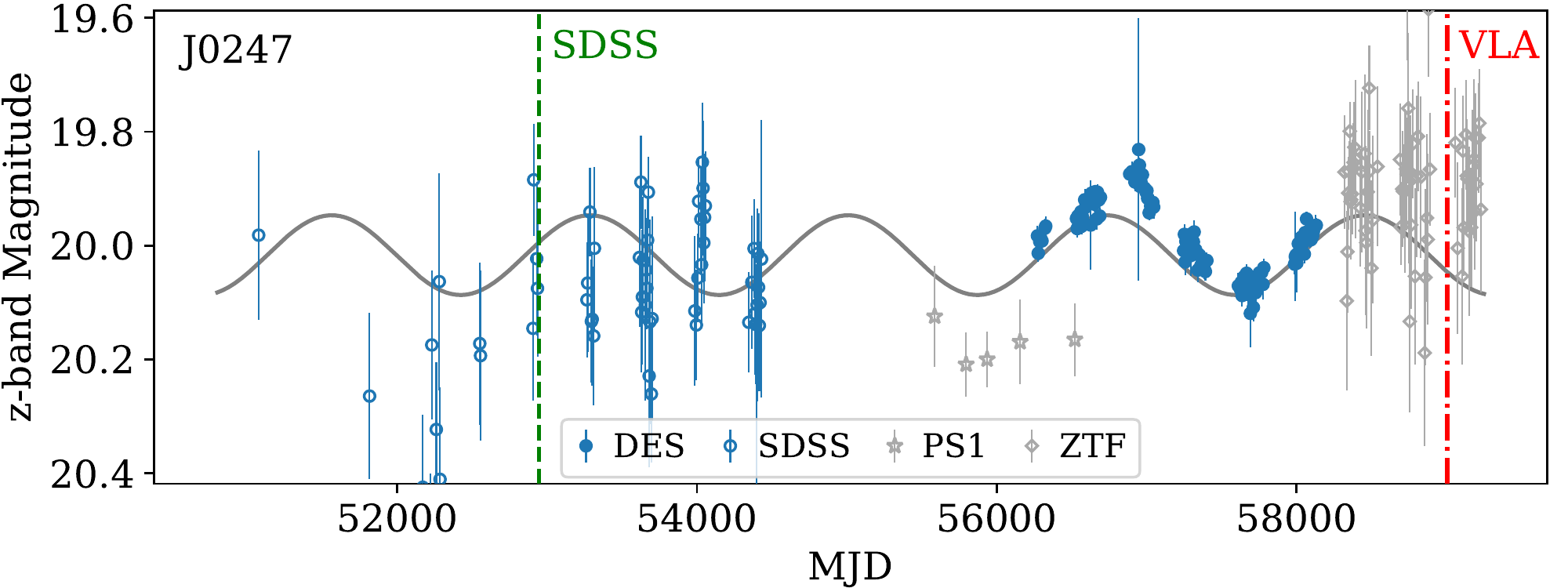}
    \includegraphics[width=\linewidth]{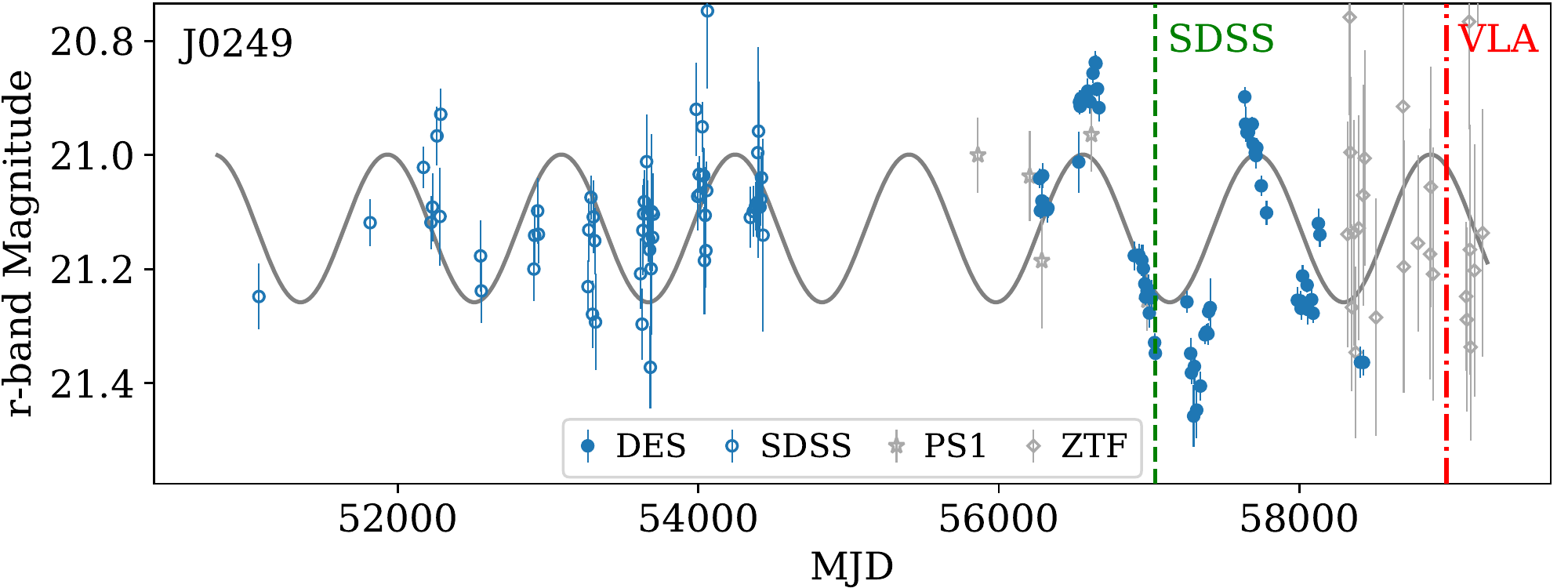}
    \includegraphics[width=\linewidth]{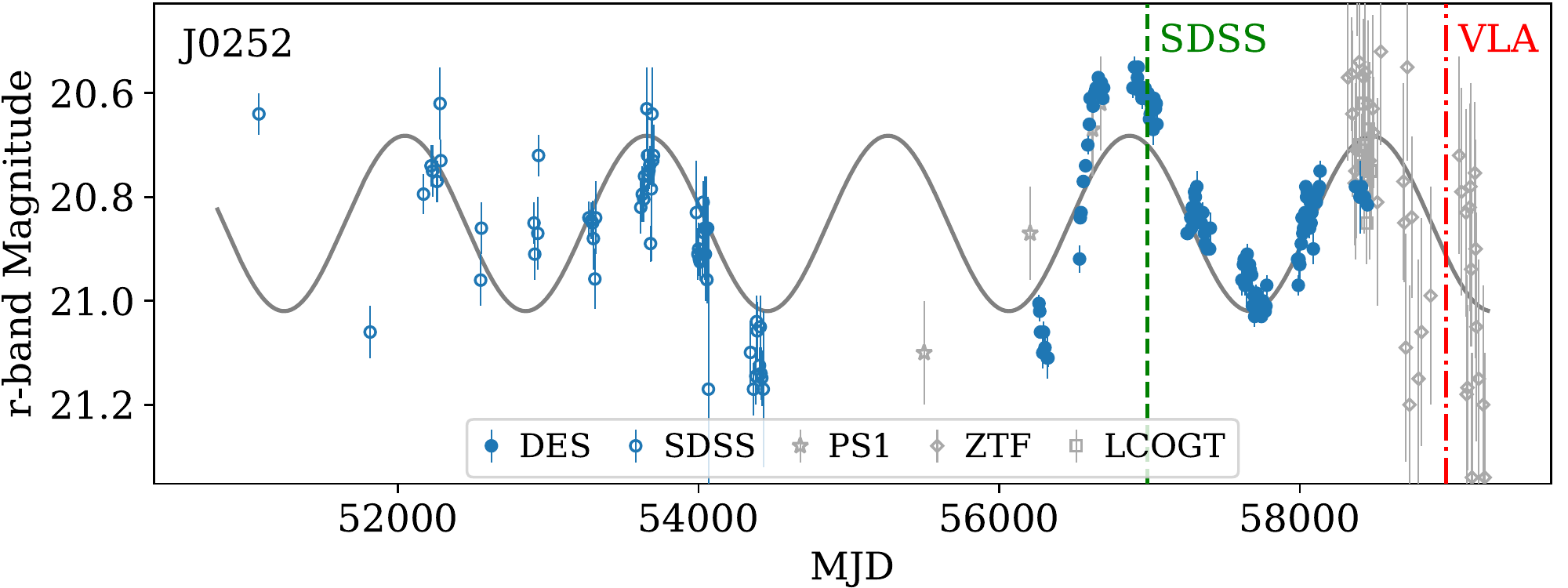}
    \caption{Optical light curves of the three DES--SDSS-selected candidate periodic quasars from \citet{chenyc2020} in the optical band containing the rest-frame 2500 {\rm \AA} ($z$-band for {\obja} and $r$-band for {\objb} and {\objc}). The ZTF $z$-band light curves of {\obja} were made from its ZTF $r$-band light curves with the color correction. 
    The grey solid curves show the best-fit sinusoidal models. The red (green) dashdot (dashed) vertical lines represent the time when the VLA (SDSS) observations were taken. The flux variations between the SDSS and the VLA observations are used to correct for the variability effect in the radio loudness measurements.}
    \label{fig:lcs}
\end{figure}

By comparing the radio continuum flux density with the optical continuum flux density at rest-frame 2500 {\rm \AA} derived from the SDSS spectra, we estimate the radio loudness parameter \citep{Kellermann1989} $R\equiv f_{6~{\rm cm}}/f_{2500}$, where $f_{6~{\rm cm}}$ is the rest-frame flux densities at 6 cm and $f_{2500}$ is the rest-frame flux densities at 2500\AA. $f_{6~{\rm cm}}$ are calculated from the VLA images assuming the radio flux follows a power law $f_{\nu}\propto v^{\alpha}$. To quantify systematic uncertainties, we adopt two different radio spectral indices $\alpha=-0.5$ \citep{Jiang2007} or $\alpha=-0.8$ \citep{Kimball2008}. $f_{2500}$ are calculated based on the archival optical SDSS spectra. Since the VLA observation and the SDSS spectra were taken at different time, we calculated the optical flux variations between the VLA and the SDSS observations using the best-fit sinusoidal models of the optical light curves shown in \autoref{fig:lcs} \citep{chenyc2020} . $f_{2500}$ has been corrected for the variability effect using the flux variation. The estimated radio loudnesses are $<$1.0, $<$1.5, and 1.2 assuming $\alpha=-0.8$ for {\obja}, {\objb} and {\objc}, respectively.
\autoref{tab:info} lists the intrinsic radio luminosities and the inferred radio loudness parameters for the three periodic quasars. 

\subsection{Multi-Wavelength Spectral Energy Distributions}

\begin{figure}
    \centering
    \includegraphics[width=\linewidth]{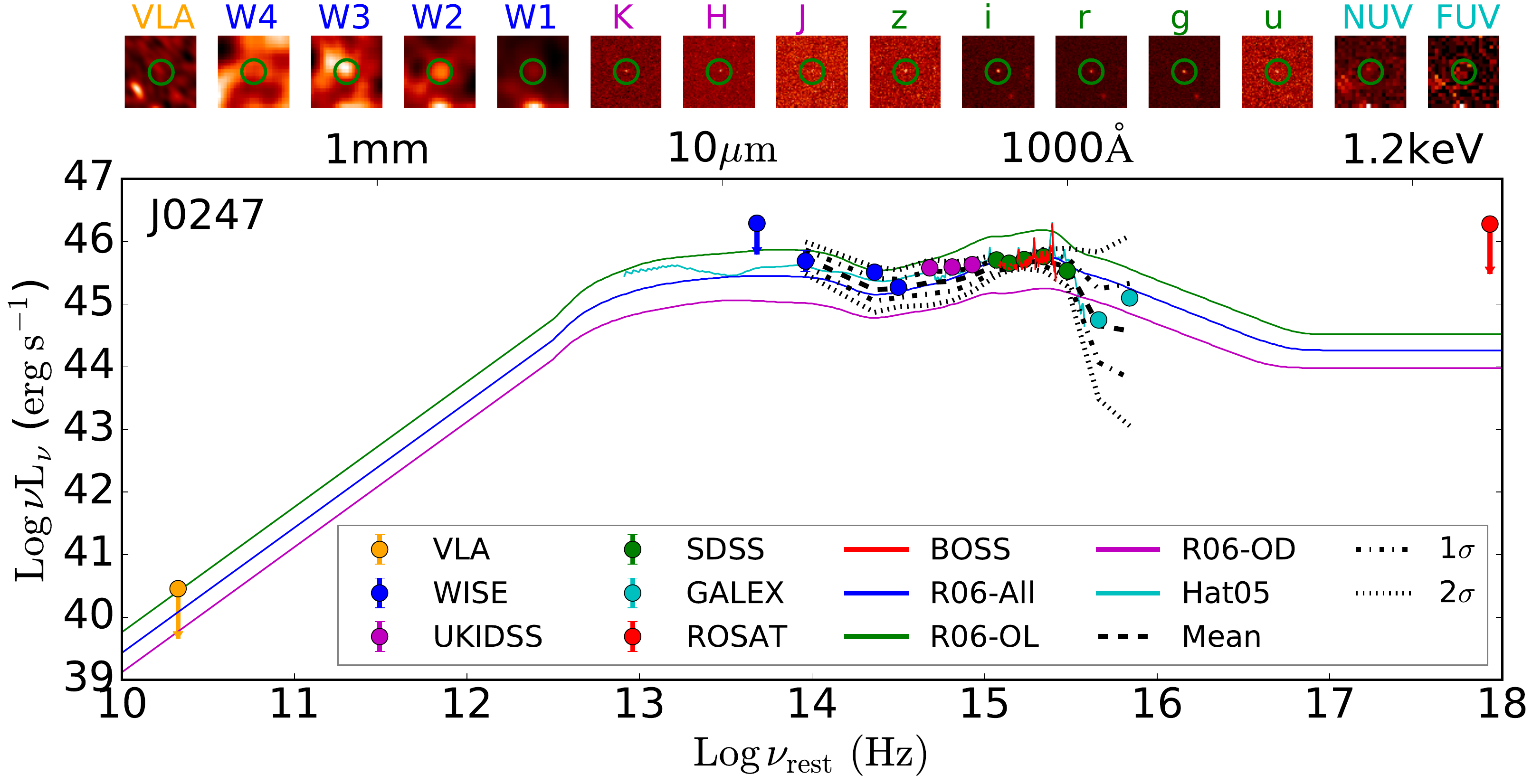}
    \includegraphics[width=\linewidth]{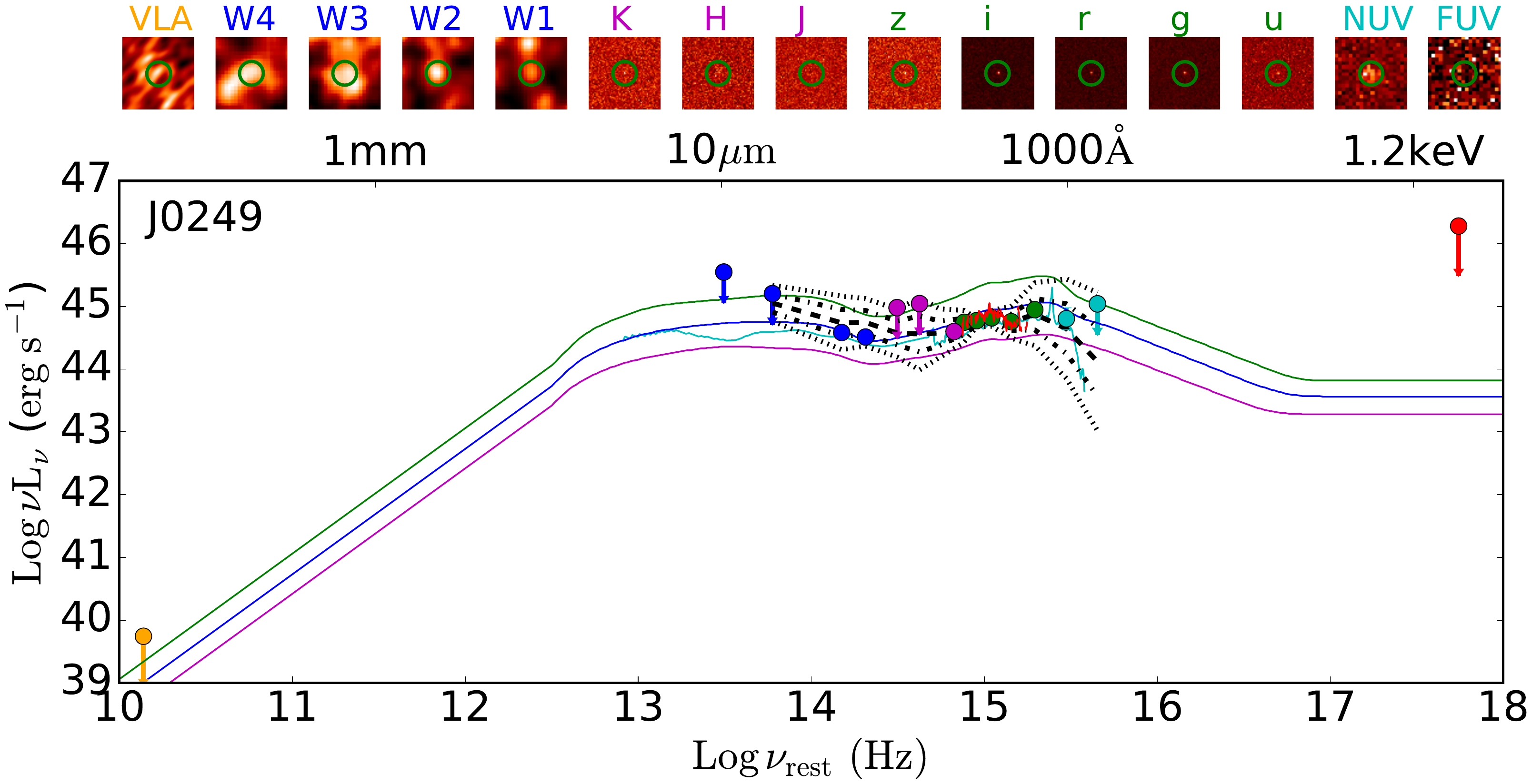}
    \includegraphics[width=\linewidth]{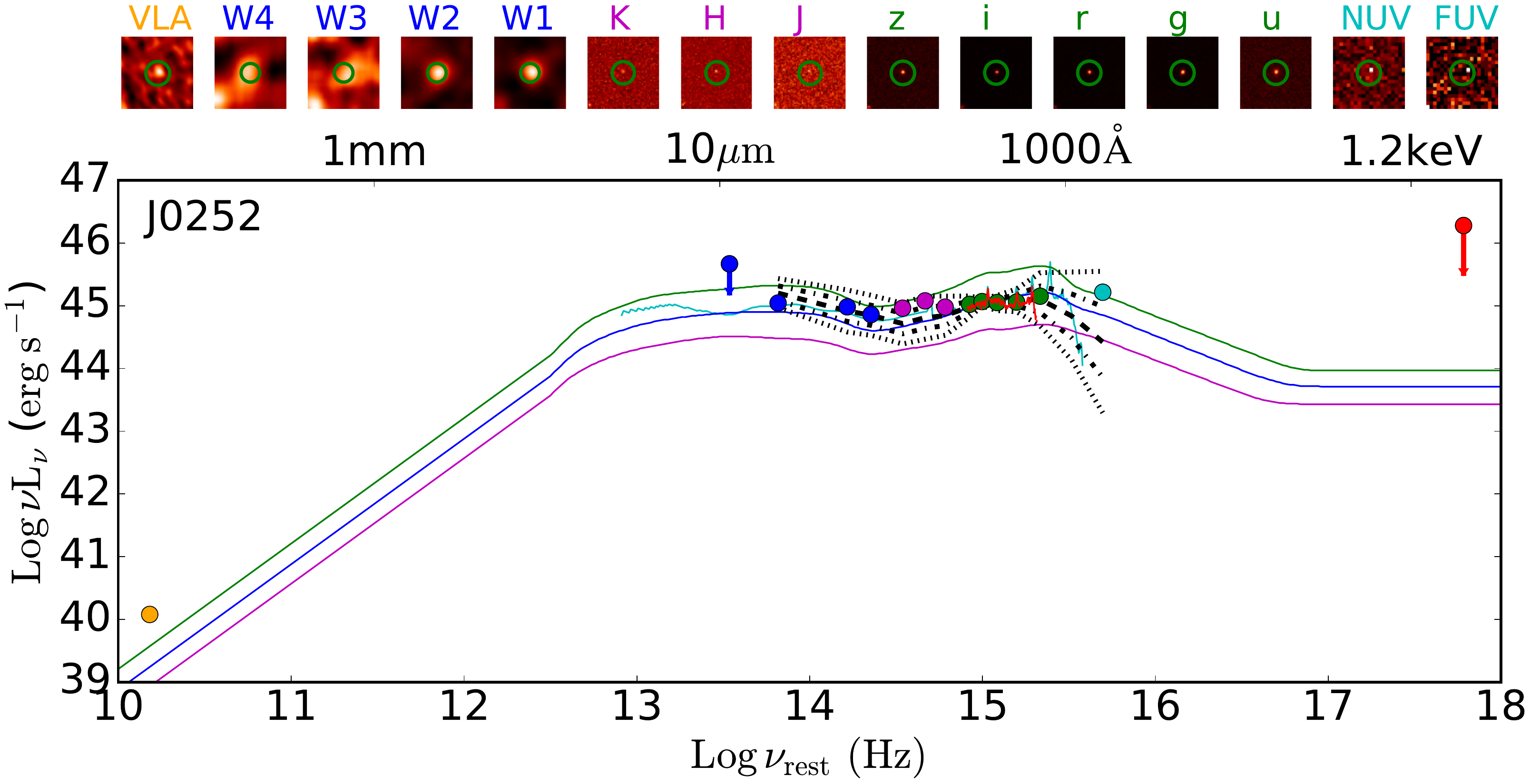}
    \caption{SEDs of the three DES--SDSS-selected candidate periodic quasars. The 3-$\sigma$ upper limits are marked as downwards arrows. The black dashed lines are the mean SEDs of the control quasar sample matched in redshift and absolute magnitude. The dashdot (dotted) black lines are the 1$\sigma$ (2$\sigma$) interval of the SEDs of the control samples. The mean SED of the whole sample in \citet{Richards2006} (``R06-All'' for all quasars, ``R06-OL'' for optically luminous quasars, and ``R06-OD'' for optically dim quasars), and the mean SED from \citet{Hatziminaoglou2005} (Hat05) are also shown as comparison. The 30\arcsec$\times$30\arcsec~ multi-wavelength images with green circles centered at the target position are displayed on top of the SEDs.}
    \label{fig:seds}
\end{figure}

\autoref{fig:seds} shows the spectral energy distributions (SEDs) of the three DES--SDSS-selected candidate periodic quasars. The SED data include our new 6 GHz radio continuum measurements from the VLA and archival mid-infrared photometry from the Wide-field Infrared Survey Explorer \citep[WISE,][]{Wright2010}, near-infrared photometry from the UKIRT Infrared Deep Sky Survey \citep[UKIDSS,][]{Lawrence2007}, optical photometry and spectra from the SDSS \citep{York2000}, UV photometry from the Galaxy Evolution Explorer \citep[GALEX,][]{Martin2005}, and X-ray 3$\sigma$ upper limits from the ROentgen SATellite \citep[ROSAT,][]{voges00}. We compared the SEDs of the three DES--SDSS-selected candidate periodic quasars with the mean SEDs of the control sample matched in redshift and absolute magnitude as well as the mean SEDs of the whole sample in \citet{Hatziminaoglou2005} and \citet{Richards2006}. The SEDs of the three DES--SDSS-selected candidate periodic quasars are consistent (within 3-$\sigma$) with the comparison SEDs.

\begin{table}
 \caption{
 Radio properties of the three DES--SDSS-selected candidate periodic quasars. Listed are the observed radio flux densities at 6 GHz, the $k$-corrected luminosities at rest-frame 6 GHz, and the radio loudness parameters $R\equiv f_{6~{\rm cm}}/f_{2500}$ assuming a radio spectral index of $\alpha=-0.5$ or $\alpha=-0.8$.
 }
 \label{tab:info}
  \addtolength{\tabcolsep}{-1pt}
 \begin{tabular}{ccccc}

  \hline\hline
 Abbreviated Name & $f_\nu$ &  $\nu L_\nu$   & $R_{\alpha=-0.5}$ & $R_{\alpha=-0.8}$ \\
 & [$\mu$Jy] & [10$^{39}$ erg s$^{-1}$] &  & \\
  \hline
  \obja & <9 & <21.4 & <0.6 & <1.0\\
  \objb & <9 & <4.5 & <1.1 & <1.5\\
  \objc & 13$\pm$3 & 9.7$\pm$2.2 & 0.8$\pm$0.2 & 1.2$\pm$0.3\\
  \hline
  \end{tabular}
\end{table}


\section{Discussions}
\label{sec:discussions}

\subsection{Origins of the Radio Emission}

The radio emission of quasars could be due to different physical mechanisms such as collimated jets, uncollimated winds, star formation activity from host galaxies, free-free emissions from photoionized gas, and/or accretion disk coronal activity \citep{Panessa2019}. The inferred radio luminosities of $\lesssim$10$^{40}$ erg s$^{-1}$ the and radio loudness parameters of order unity place the three DES--SDSS-selected candidate periodic quasars in the radio-quiet regime \citep{Kellermann1989}.

\citet{Zakamska2014} analyzed 568 obscured luminous quasars with median radio luminosity $\nu L_\nu$ [1.4 GHz] of $10^{40}$ erg s$^{-1}$ and concluded that the radio emissions in radio-quiet quasars are due to uncollimated outflows. The 6 GHz study in a volume-limited sample of 178 low-redshift (0.2$<z<$0.3) quasars by \citet{Kellermann2016} suggests that the radio emission in 80\% of radio-quiet quasars is powered by star formation in their host galaxies. According to the luminosity function shown in Figure 9 in \citet{Kellermann2016}, the radio luminosities of the DES--SDSS-selected candidate periodic quasars are likely powered by star formation or a mixture of uncollimated quasar outflows and star formation. More recently, \citet{Nyland2020} discovered compact young jets in quasars that had recently brightened in the radio. These authors suggest that young and short-lived jets which do not grow to large scales might be common at high redshift. However, the radio fluxes of these young-jet-powering quasars are ten to a hundred times higher than those of our targets. 

In summary, the relatively weak radio luminosities of the DES--SDSS-selected candidate periodic quasars are unlikely from collimated radio jets. Even if any weak, small-scale jets exist, which would have been unresolved in the VLA images given the few arcsecond resolution, the large-amplitude ($\gtrsim$20\%) variability detected in the optical light curves could not be explained by pure optical contribution from precessing radio jets based on the relative weaknesses of the radio luminosities.

\subsection{Comparison to Previous Candidate Periodic Quasars}


OJ287 and PG 1302$-$102 are two prototypical candidate periodic quasars that have been proposed as candidates for binary SMBHs \citep{Lehto1996,valtonen08,Graham2015a}. OJ287 is a low-redshift ($z = 0.306$) blazar \citep{Stickel1989} showing candidate double-peaks in its optical light curves every 12 years \citep[][but see \citealt{Goyal2018} for evidence against any significant periodicity]{Sillanpaa1988,Lehto1996}. The periodic peaks since 1891 have been interpreted as arising from the binary motions at similar timescale. OJ287 was detected by the {\it Swift} satellite in the X-rays \citep{DElia2013} and by the Fermi Large Area Telescope in the $\gamma$-rays \citep{Fermi_cat4}.  PG1302$-$102 is also a low-redshift ($z = 0.278$) blazar \citep{Massaro2015}. It was detected by the {\it Swift} satellite in the X-rays \citep{Swift70}. 
PG 1302$-$102 shows a 5.2-yr periodic signal with 1.5 cycles covered by the Catalina Real-Time survey \citep[][see also \citealt{Liutt2018,Zhu2020}]{Graham2015a}. Doppler boosting has been proposed to explain the optical light curve periodicity \citep{DOrazio2015,Xin2020a,Song2021}. 

Unlike our targets of DES--SDSS-selected candidate periodic quasars, OJ287 and PG 1302$-$102 have strong radio emission with flux densities on the order of hundreds of mJy. They share a similar core-jet structure as seen in both the optical and radio observations \citep{Hutchings1994,Benitez1996}. In contrast, the three DES--SDSS-selected candidate periodic quasars only show weak radio cores with flux densities of $\lesssim$ tens $\mu$Jy and no associated jet components. The comparison between the radio and the optical images suggests that their morphologies are consistent with unresolved point sources. There is a faint emission 2{\arcsec} N-E of {\objb}, although no radio emission is seen at the same position, indicating that it might be a neighboring galaxy, or from the superposition of a foreground star or galaxy.

\begin{figure}
    \centering
    \includegraphics[width=\linewidth]{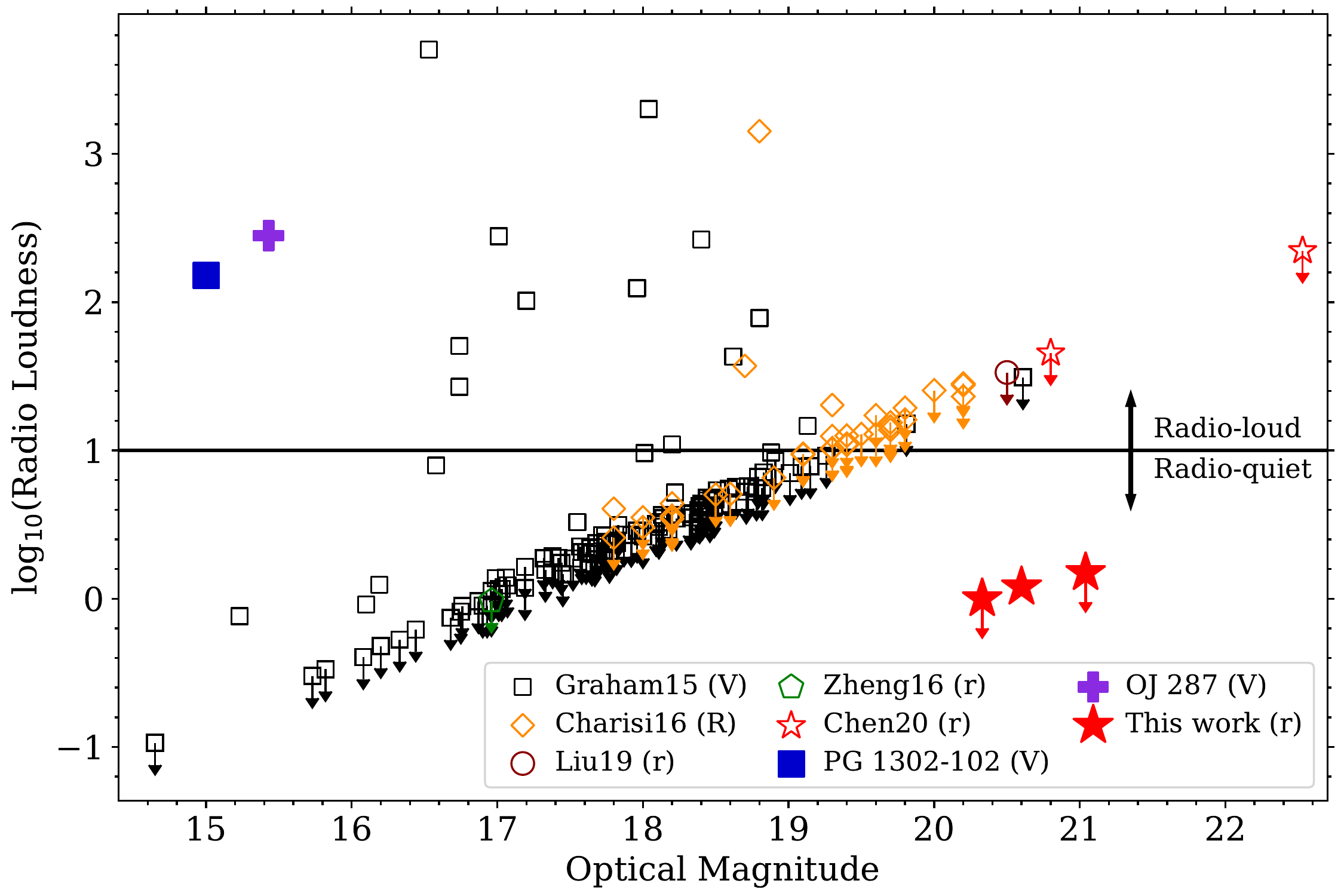}
    \caption{Radio Loudnesses of the periodic quasars from previous searches \citep{Graham2015,Charisi2016,Zheng2016,Liutt2019,chenyc2020} and two well-known periodic quasars PG1302-102 and OJ287. The points with downward arrows represent the  upper limits estimated from the detection limits of 1 mJy in FIRST \citep{FIRST}. The solid line marks the $R$=10 boundary for distinction between radio-loud and radio-quiet quasars \citep{Kellermann1989}.
    }
    \label{fig:mag_RL}
\end{figure}

\autoref{fig:mag_RL} compares the radio loudnesses $R$ of the DES--SDSS-selected candidate periodic quasar with those of other candidate periodic quasars from the literature \citep{Graham2015,Charisi2016,Liutt2019,Zheng2016,chenyc2020} including PG1302$-$102 and OJ287. For the comparison quasars, we follow the similar procedures of calculating the radio loudness as those described in \autoref{sec:results-RL}. The optical flux densities are estimated from broad-band photometry and are scaled using the frequency power-law index, $\alpha_{\nu,{\rm opt}}$ of $-0.44$ \citep{VandenBerk2001}. The radio flux densities are estimated from the VLA sky survey \citep[VLASS,][]{VLASS} and the Faint Images of the Radio Sky at Twenty-cm \citep[FIRST,][]{FIRST}. The radio loudness of the three DES--SDSS-selected candidate periodic quasars are hundreds times smaller than those of PG1302-102 and OJ287. For the other candidate periodic quasars in the literature, which are mostly from the CRTS and the PTF, $\sim$80\% of them are undetected in FIRST and/or VLASS due to the limited sensitivity, which makes the comparison incomplete especially for faint objects. Nevertheless, we can still put a lower limit of $\sim$10\% for the other candidate periodic quasars from the literature which are radio-loud ($R>$10).

\citet{Guo2020} has carried out a multi-wavelength SED study for the previous candidate periodic quasars from the CRTS and the PTF. These authors find a tentatively higher fraction of radio-loud quasars compared to the control sample of ordinary quasars matched in redshift and luminosity. This has been interpreted as possible contamination from processing radio jets, although the sample is subject to false positives due to the limited cycles of periodicities being observed  \citep{Vaughan2016}. The higher fraction of radio-loud quasars could be due to a selection bias given that blazars exhibit much larger variability amplitudes. Due to the faintness of the DES--SDSS-selected candidate periodic quasars, the control sample of ordinary quasars matched in redshift and luminosity are mostly undetected in the current archival radio surveys including VLASS and FIRST. It is unsurprising that the control sample of ordinary quasars are undetected in the wide-field radio surveys since 80\% of the quasars are radio-quiet \citep{Kellermann2016}. 


\subsection{Implications for the Radio Jet Precession Scenario}

Jet precession is one of the popular scenarios invoked to explain the radio periodicity observed in some radio-loud quasars \citep{Bach2006,Britzen2010,Kudryavtseva2011,Kulkarni2016}. With a powerful radio jet, the synchrotron emission could contribute significantly to the UV/optical emission \citep{Crane1993,Sparks1995,Scarpa1999}. For blazars with observed optical periodicities, while a binary SMBH is a plausible explanation \citep{Lehto1996,Villata1998,Valtaoja2000,Caproni2017,Sobacchi2017}, the scenario of a precessing radio jet powered by a single SMBH remains possible \citep{Rieger2004,Britzen2018,Butuzova2020}. The precession of a helical jet with a small angle between the helix axis and the line of sight could explain the periodic bursty light curves.

The jet precession scenario either from a binary SMBH or a single SMBH, however, is highly unlikely for the three DES--SDSS-selected candidate periodic quasars studied here. The radio-quiet properties suggest that the optical periodicity is not being significantly contaminated with radio jets. The optical spectra with typical quasar-like power-law continuum and strong emission lines \citep{chenyc2020} also distinguish them from the featureless systems such as BL Lacs, whose emission is dominated by jets. Therefore, we anticipate that the optical periodicity is associated with the accretion disk. Different physical origins of the optical periodicity for the DES--SDSS-selected periodic quasars including Doppler boosting, hydrodynamic variability in circumbinary disk, or a warped accretion disk have been discussed in \citet{chenyc2020}. It has been shown that Doppler boosting and a warped accretion disk are difficult to explain the characteristics of the optical light curves  \citep[multi-wavelength dependence, processing and damping timescale, and variability amplitude; see][]{chenyc2020}. \citet{liao2020} found that {\objc} is best explained by a bursty model predicted by hydrodynamic simulations of circumbinary accretion disks. There are other scenarios like shorter time variability attributed to nutation on top of a longer precession period or quasi-periodic behavior due to instability in accretion flow, but detailed light curve modeling to test them is beyond the scope of this work.

The DES--SDSS-selected candidate periodic quasars are at higher redshift ($z\sim2$) than those from previous samples of candidate periodic quasars and might represent a new radio-quiet population different from those low-redshift periodic blazars whose emissions are dominated by synchrotron radiation. This higher-redshift, radio-quiet population is powered by less massive SMBHs \citep{chenyc2020} whose mergers are more relevant for future space-born gravitational wave experiments such as the Laser Interferometer Space Antenna \citep[e.g.,][]{HolleyBockelmann2020,Volonteri2020} and TianQin \citep[e.g.,][]{WangHT2019}. Future sensitive synoptic surveys such as LSST \citep{Ivezic2019} will be able to detect more periodic quasar candidates similar to the DES--SDSS-selected candidate periodic quasars \citep{Xin2021}. Together with deep multi-wavelength follow-ups on those periodic quasars, we will be able to characterize the population in detail and systematically test different hypotheses to explain their optical periodicities.


\section{Conclusions}
\label{sec:conclusions}

We have presented deep 6.0 GHz VLA images for the three DES--SDSS-selected candidate periodic quasars discovered by \citet{chenyc2020}. We have measured their radio loudness parameters and analyzed their radio properties as compared against those of other candidate periodic quasars from the literature. We summarize our main findings and as follows:
\begin{enumerate}
\item The $k$-corrected intrinsic radio luminosities at rest-frame 6 GHz are $<2.1\times10^{40}$, $<4.5\times10^{39}$ and ${\sim}9.7\times10^{39}$ erg s$^{-1}$ and the radio loudnesses are $<$1.0, $<$1.5 and ${\sim}$1.2 for {\obja}, {\objb} and {\objc}, respectively. 

\item The radio luminosity of $\lesssim10^{40}$ erg s$^{-1}$ and the radio loudness of order unity place the three candidate periodic quasars in the radio-quiet regime, where significant contamination in the optical emission from radio jets is highly unlikely. The origin of the radio emission is likely from star formation and/or uncollimated quasar winds. Together with the core-only morphology in the radio images, our results rule out the radio jet procession as the origin of the observed optical periodicity. The optical periodicity is likely associated with the variability in accretion disk.

\item Unlike the previously known periodic blazars such as OJ 287 and PG 1302$-$102, the DES--SDSS-selected candidate periodic quasars at $z\sim2$ show radio-quiet properties, suggesting that they might represent a different population from those found in low-redshift radio-loud quasars.   

\end{enumerate}

The VLA observations of the three DES--SDSS-selected candidate periodic quasars serve as a pilot study of radio properties for periodic quasars. Future Legacy Survey of Space and Time \citep[LSST;][]{Ivezic2019} of the Vera C. Rubin Observatory will discover $\gtrsim$hundreds of candidates similar to the DES--SDSS-selected candidate periodic quasars. Deep multi-wavelength observations will be needed to characterize them in detail and systematically evaluate the competing hypotheses for the optical periodicity before they can be established as binary SMBHs.

\section*{Acknowledgements}

We thank the anonymous referee for the constructive comments and suggestions. We thank Amy Mioduszewski, Anthony Perreault and Heidi Medlin for help with our VLA observations. Y.-C.C. and X.L. acknowledge a Center for Advanced Study Beckman fellowship and support from the University of Illinois campus research board. Y.-C.C. is supported by the Illinois Graduate Survey Science Fellowship and the government scholarship to study aboard from the ministry of education of Taiwan. H.G. acknowledges support from National Science Foundation grant AST-1907290

The National Radio Astronomy Observatory is a facility of the National Science Foundation operated under cooperative agreement by Associated Universities, Inc.

This project used public archival data from the Dark Energy Survey (DES). Funding for the DES Projects has been provided by the U.S. Department of Energy, the U.S. National Science Foundation, the Ministry of Science and Education of Spain, the Science and Technology FacilitiesCouncil of the United Kingdom, the Higher Education Funding Council for England, the National Center for Supercomputing Applications at the University of Illinois at Urbana-Champaign, the Kavli Institute of Cosmological Physics at the University of Chicago, the Center for Cosmology and Astro-Particle Physics at the Ohio State University, the Mitchell Institute for Fundamental Physics and Astronomy at Texas A\&M University, Financiadora de Estudos e Projetos, Funda{\c c}{\~a}o Carlos Chagas Filho de Amparo {\`a} Pesquisa do Estado do Rio de Janeiro, Conselho Nacional de Desenvolvimento Cient{\'i}fico e Tecnol{\'o}gico and the Minist{\'e}rio da Ci{\^e}ncia, Tecnologia e Inova{\c c}{\~a}o, the Deutsche Forschungsgemeinschaft, and the Collaborating Institutions in the Dark Energy Survey.
The Collaborating Institutions are Argonne National Laboratory, the University of California at Santa Cruz, the University of Cambridge, Centro de Investigaciones Energ{\'e}ticas, Medioambientales y Tecnol{\'o}gicas-Madrid, the University of Chicago, University College London, the DES-Brazil Consortium, the University of Edinburgh, the Eidgen{\"o}ssische Technische Hochschule (ETH) Z{\"u}rich,  Fermi National Accelerator Laboratory, the University of Illinois at Urbana-Champaign, the Institut de Ci{\`e}ncies de l'Espai (IEEC/CSIC), the Institut de F{\'i}sica d'Altes Energies, Lawrence Berkeley National Laboratory, the Ludwig-Maximilians Universit{\"a}t M{\"u}nchen and the associated Excellence Cluster Universe, the University of Michigan, the National Optical Astronomy Observatory, the University of Nottingham, The Ohio State University, the OzDES Membership Consortium, the University of Pennsylvania, the University of Portsmouth, SLAC National Accelerator Laboratory, Stanford University, the University of Sussex, and Texas A\&M University.
Based in part on observations at Cerro Tololo Inter-American Observatory, National Optical Astronomy Observatory, which is operated by the Association of Universities for Research in Astronomy (AURA) under a cooperative agreement with the National Science Foundation.

Funding for the Sloan Digital Sky 
Survey IV has been provided by the 
Alfred P. Sloan Foundation, the U.S. 
Department of Energy Office of 
Science, and the Participating 
Institutions. 

SDSS-IV acknowledges support and 
resources from the Center for High 
Performance Computing  at the 
University of Utah. The SDSS 
website is www.sdss.org.

SDSS-IV is managed by the 
Astrophysical Research Consortium 
for the Participating Institutions 
of the SDSS Collaboration including 
the Brazilian Participation Group, 
the Carnegie Institution for Science, 
Carnegie Mellon University, Center for 
Astrophysics | Harvard \& 
Smithsonian, the Chilean Participation 
Group, the French Participation Group, 
Instituto de Astrof\'isica de 
Canarias, The Johns Hopkins 
University, Kavli Institute for the 
Physics and Mathematics of the 
Universe (IPMU) / University of 
Tokyo, the Korean Participation Group, 
Lawrence Berkeley National Laboratory, 
Leibniz Institut f\"ur Astrophysik 
Potsdam (AIP),  Max-Planck-Institut 
f\"ur Astronomie (MPIA Heidelberg), 
Max-Planck-Institut f\"ur 
Astrophysik (MPA Garching), 
Max-Planck-Institut f\"ur 
Extraterrestrische Physik (MPE), 
National Astronomical Observatories of 
China, New Mexico State University, 
New York University, University of 
Notre Dame, Observat\'ario 
Nacional / MCTI, The Ohio State 
University, Pennsylvania State 
University, Shanghai 
Astronomical Observatory, United 
Kingdom Participation Group, 
Universidad Nacional Aut\'onoma 
de M\'exico, University of Arizona, 
University of Colorado Boulder, 
University of Oxford, University of 
Portsmouth, University of Utah, 
University of Virginia, University 
of Washington, University of 
Wisconsin, Vanderbilt University, 
and Yale University.

Based on observations obtained with the Samuel Oschin 48-inch Telescope at the Palomar Observatory as part of the Zwicky Transient Facility project. ZTF is supported by the National Science Foundation under Grant No. AST-1440341 and a collaboration including Caltech, IPAC, the Weizmann Institute for Science, the Oskar Klein Center at Stockholm University, the University of Maryland, the University of Washington, Deutsches Elektronen-Synchrotron and Humboldt University, Los Alamos National Laboratories, the TANGO Consortium of Taiwan, the University of Wisconsin at Milwaukee, and Lawrence Berkeley National Laboratories. Operations are conducted by COO, IPAC, and UW.

{\it Facilities:} VLA, DES, Sloan

{\it Software:} CASA \citep{McMullin2007}, PyQSOFit \citep{Guo2018,Shen2019}, Astropy \citep{astropy:2013, astropy:2018}



\section*{Data Availability}

The VLA data will be available in the VLA data archive at \url{https://science.nrao.edu/facilities/vla/archive/index} after the proprietary period. The DES data are available at \url{https://des.ncsa.illinois.edu}. The SDSS spectra are available in SDSS Data Release 16 at \url{https://www.sdss.org/dr16}.



\bibliographystyle{mnras}
\bibliography{reference} 




\appendix


\bsp	
\label{lastpage}
\end{document}